\documentstyle[12pt,aps,epsf]{revtex}

\input{epsf}
\begin{document}

\title{Saturable Absorber, Coherent Population Oscillations,
\vskip 1mm and "Slow Light"}
\author{V.S.Zapasskii and G.G.Kozlov}

\maketitle
 \hskip20pt
\vskip20pt
\hskip100pt {\it e}-mail:  gkozlov@photonics.phys.spbu.ru
\vskip20pt
All-Russia Research Center "Vavilov State Optical Institute",
St. Petersburg, 199034 Russia
\begin{abstract}
The paper presents a critical analysis of publications
devoted to one of the methods for production of the so-called
"slow light" (the light with an anomalously low group velocity)
due to extremely high local steepness of the dispersion curve of the medium. The
method in point employs for this purpose the effect of coherent
population oscillations accompanied by burning of a narrow
spectral hole in the homogeneously broadened absorption spectrum. Physical model of the effect proposed in the studies
under consideration is based on analysis of response of a
nonlinear medium to a low-frequency intensity modulation of
the incident light beam. We show that all the observations
described in these papers can be easily interpreted in the
framework of the simplest model of saturable absorber and have
nothing to do with the effects of hole burning and group velocity
reduction.
\end{abstract}
\newpage
\section{INTRODUCTION}

 After publication of the first impressive results on group
velocity reduction by several orders of magnitude \cite{1,2}, the
effects of slowing down, stoppage, storage, and release of light
attracted attention of a great number of researchers. The
experiment \cite{2}, which demonstrated the light velocity reduction
by more than 7 orders of magnitude, was performed with
sodium atoms at the temperatures close to that of the Bose-Einstein condensation (hundreds of nK) and was based on the
effect of electromagnetically induced transparency. The paper \cite{2} initiated a great
number of publications, with some of them, in particular,
claiming observation of the light velocity reduction of similar
scale under much less exotic experimental conditions (e.g., at
room temperature). To create a steep dispersion of the refractive index,
along with the effect of electromagnetically induced
transparency, there has been used the effect of coherent
population oscillations (CPO). Observation of the CPO-based
hole burning in the absorption spectra of ruby and
alexandrite crystals was reported in  \cite{3,4}. Then, in \cite{5,6}, there has been
announced observation, in these crystals, of "slow" and "fast" light, i.e., the light with "ultralow" and, in this particular
case, "negative" group velocity. More recently, other
publications have appeared that supported the proposed
interpretation \cite{7,8}, and, as a result, the CPO effect was included into the
arsenal of the "slow light" methods (see, e.g.,
\cite{9,10}).

The goal of this publication is to show that the light-pulse delay
detected in the above experiments is a well known effect of nonlinear
optics. It has nothing in common with the CPO effect
and is not connected with the group velocity reduction
in the medium. In our opinion, the experimental data of \cite{3,4,5,6} provide no evidence for the "slow light" observation, and their interpretation is erroneous.

\section{THE EFFECT OF COHERENT POPULATION
OSCILLATIONS}

The role of population oscillations in the coupling of the light
waves with close frequencies propagating in a resonant
nonlinear medium was first noted, as far as we know, by W.
Lamb in 1964 \cite{11}. Analysis of the properties of a saturable absorber
performed by Schwartz and Tan in 1967 \cite{12} has shown that, in
the presence of a strong (saturating) pump wave, a weak probe
wave scanned in the vicinity of the strong wave frequency
displays an apparent dip in the absorption spectrum with the center at
the pump frequency and with the width
corresponding to the inverse absorption relaxation time. The
effect had a simple and visual physical interpretation. The
combined action of two monochromatic waves, due to their
interference, gives rise to a modulation of the total field intensity  at
the beat frequency. When this frequency appears to be
sufficiently low (comparable with the inverse population
relaxation time or lower), the light intensity modulation is
transformed into modulation of populations (and absorption) of
the optically active centers. Under these conditions, interaction
of the monochromatic pump wave with the oscillating
susceptibility of the medium destroys its monochromaticity and
gives birth to sidebands, with one of them being exactly
coincident in frequency with the probe wave and thus contributing to
its intensity. This is revealed as an apparent bleaching of the
medium (or as a dip in the absorption spectrum) in the vicinity
of the pump frequency. The width of such a dip is, as a rule,
smaller than homogeneous width of the optical transition. For
this reason, this effect was often considered as "hole burning" in
a homogeneously broadened spectrum.

The interest to the CPO effect in 70s of the last century was
stimulated, on the one hand, by its role in the laser emission
dynamics and, on the other, by the possibility to use it for
getting information about lifetimes of excited states in
nonluminescent or rapidly relaxing systems. In those years, the CPO effect
has been thoroughly studied both theoretically (see, e.g.,
\cite{13,14}) and experimentally \cite{15,16}. The experimental studies
mainly demonstrated applicability of this spectroscopic phenomenon  to
kinetic measurements. In spite of the fact that this effect is linear with respect to the probe wave and is phenomenologically close to the well-known effect of hole-burning in
the inhomogeneously broadened spectra, it results from the multiwave interaction and is, evidently, of essentially different nature.

The conclusion about observation of a narrow spectral dip ($\Delta\nu \approx 37$
Hz) in the homogeneously broadened absorption
spectrum of ruby crystal was made in \cite{3} based on analysis
of the crystal's response to the low-frequency intensity
modulation of the light beam (Ar$^+$-laser). The dip was
treated as a usual spectral feature in the complex refractive
index of the medium. In \cite{5}, to this narrow spectral dip, in
conformity with the Kramers-Kronig relations, was assigned an
extremely steep slope of the dispersion curve, and  the
observation of "slow light" with a group velocity of 57.5 $\pm$
0.5 m/s was reported. Similar experiments performed with
alexandrite crystals were reported in \cite{4,6}. This crystal,
behaved, at certain wavelengths, as an {\it inverse} saturable
absorber (with the absorption growing with light intensity), and the
CPO effect, in this crystal, resulted in "anti-hole burning" in
the absorption spectrum. The experimental observations of \cite{6} have
been interpreted in terms of "negative" group velocity of light
("fast" light). These papers provided the basis for a new
approach to the problem of implementation of the
"slow" and "fast" light. To evaluate the novelty and validity of
this interpretation, let us turn to basic properties of a saturable
absorber.

\section{SATURABLE ABSORBER}

Characteristic properties of a saturable absorber attracted attention of
researchers in the early laser epoch, in 60s, when the
phthalocyanine dyes were found to be efficient for laser Q-switching. In
those years, their temporal and intensity-related characteristics
have been studied in sufficient detail \cite{17,18,19,20}. The simplest
saturable absorber is known to be well modeled by a two-level
system with a high dephasing rate $T_2^{-1} (T_2^{-1} >> T_1^{-1}, \Omega$,
 where $T_1$ is the population relaxation time and
$\Omega$ is the Rabi frequency). As a good example of the
saturable absorber may also serve ruby crystal (Fig. 1a). The
energy structure of chromium ion in ruby is not two-level, but a
high relaxation rate of the excited state $^4F_2$  (($|b\rangle$ in Fig. 1) to
the metastable levels $2\bar{A}$ and $\bar{E}$ ($|c\rangle$) makes it possible to
neglect the coherent transient processes practically at any
reasonable excitation power.

In the general case (neglecting the effects of propagation),
dependence of intensity of the light transmitted through a
saturable absorber $I_{out}$ on its intensity at the entrance $I$
can be presented in the form
\begin{equation}
I_{out}= \ae(I,t)I,
\end{equation}

where the transmissivity factor $\ae(I,t)$ controls all the dynamic
and intensity-related characteristics of the saturable absorber.

Under nonstationary conditions, when the light intensity varies
in time, the dynamics of transmission of the saturable absorber
is dedescribed by the equation

 \begin{equation}
 \dot{\ae}={\ae_{eq}-\ae\over\tau}
 \end{equation}

Here, $\ae_{eq}$ is the equilibrium value of $\ae$
corresponding to the current light intensity, and $\tau$ is a
time constant (at low light intensities $\tau \approx T_1$. When the light intensity is not too high, the
dependence $\ae_{eq}(I)$ can
be presented in the form

\begin{equation}
\ae_{eq}(I)=\ae_0-\ae_1 I, \hskip10mm \ae_1 I<<\ae_0.
\end{equation}

As the light intensity increases, the nonlinear is usually bleached. In this case, the parameter $\ae_1$ is negative.
There are situations, however, when absorption of the optical
medium increases with increasing light intensity, rather than
decreases. This situation is typical for certain schemes of optical
orientation and is realized, in particular, in the alexandrite
crystal mentioned above, where, due to efficient excited-state
absorption, chromium ions in the excited metastable state may absorb
light, at the wavelength of excitation, stronger than those in the
ground state (Fig. 1b). For these {\it inverse}
saturable absorber, $\ae_1>0$.

Solving Eqs. (1)--(3) for a stepwise change of the incident light
intensity, one can easily make sure that the time dependence of
the transmitted light intensity, in this case, contains two
components - an inertialess (stepwise) jump and an exponential component,
reflecting dynamics of establishment of steady-state
populations of the medium. The sign of the exponential
component, evidently, depends on the sign of $\ae_1$
(Fig. 2).  In virtue of linearity of the Fourier-transform, the appropriate
response in the frequency domain
will also contain two components - a frequency-
independent ("white") pedestal and a Lorentz-wise peak in the range of
zero frequencies with a width of $\sim 1/\tau$. This dependence
can be easily obtained in analytical form by calculating the
response of the saturable absorber to the light beam with
harmonically modulated intensity

\begin{equation}
I=I_0+I_1\exp(i\omega t),
\end{equation}

>From Eqs. (1)--(3), neglecting the terms $\sim I_1^2$, we have:

\begin{equation}
I_{out}(t)=I_0(\ae_0-\ae_1 I_0)+I_1\exp(i\omega t) K(\omega)
\end{equation}
\begin{equation}
K(\omega)\equiv \ae_0-\ae_1 I_0\bigg(
1+{1\over 1+i\omega\tau}
\bigg)=|K(\omega)| e^{ i\phi}
\end{equation}

Here, as usual, the complex intensities $I$ and $I_{out}$ describe
the amplitude-phase relations between harmonically oscillating light
intensities at the entrance and at the exit of the absorber.

Figure 3 shows characteristic frequency dependences of the
amplitude $K(\omega)$, phase $\phi(\omega)$, and time delay
$\delta(\omega) = \phi(\omega)/\omega$ of intensity oscillations of the
light beam
transmitted through a usual (bleachable) saturable absorber (a)
and through an inverse saturable absorber (b).

As is seen from these curves, the greatest time delay is
experienced by low-frequency spectral components of the light
intensity ($\omega << \tau^{-1}$), for which, taking into
account the condition
 $\ae_1 I_0<<\ae_0$
 from Eq. (3),  we have

\begin{equation}
\Delta t_{\hbox{max}}=\Delta t(\omega\rightarrow 0)=-{\ae_1 I_0\over \ae_0}\tau
\end{equation}

This formula yields the time delay experienced by a smooth
pulse with the width much larger than the relaxation tine
$\tau$. As expected, the time delay may constitute only a small
fraction of the pulse width.

Figure 4 shows normalized light pulses at the entrance (solid
lines) and at the exit (dashed lines) of the usual (a) and inverse
(b) saturable absorber for several ratios of the pulse width
$\delta$ and relaxation time $\tau$. As seen from the figure,
the pulse propagating through a bleachable absorber generally
experiences positive time delay, whereas in an inverse absorber
the delay is negative. Note once again that all these results have been known
for several decades (see, e.g., paper \cite{18}, in which the shape of
the pulse transmitted through a saturable absorber is calculated).
Let us now turn back to the experiments on observation of "slow
light" under conditions of coherent population oscillations \cite{5,6}.

\section{DISCUSSION}

Comparison of the results presented above with the
experimental data of \cite{3,4,5,6} did not allow us to find any
noticeable distinctions between them. Specifically, in full
agreement with the model of saturable absorber, the amplitude
of the light intensity modulation at the exit of the absorber
shows a Lorentz-wise peak in the region of low frequencies
$\omega \lesssim 1/\tau$
 (Fig. 3 from \cite{3}), and the phase shift of
the intensity modulation, at these frequencies, is positive or
negative depending on the type of the absorber (Fig. 3 from \cite{5}
and Fig. 2 from \cite{6}). A positive time delay is experienced also
by Gaussian pulse transmitted through a bleachable absorber
(cf. Fig. 4 from \cite{5} and Fig. 4 of the present paper).

Thus, all the observations of \cite{3,4,5,6}, including "slow" and "fast"
light, can be interpreted in a fairly trivial way, in the framework
of the simplest model of saturable absorber. This model,
evidently, does not imply either burning of a narrow spectral
hole or any modification of the light group velocity in the
medium.

The authors of \cite{3,4,5,6} interpret their experimental results in a
more complicated way by representing the harmonically
modulated light beam in the form of three spectral components,
with the central one playing the role of the pump and the
sidebands playing the role of the probe. Such an approach was
has been repeatedly used to describe interaction of a modulated
monochromatic light with a saturable absorber and is, in
principle, quite correct \cite{14,21}. First of all, however, the authors
of \cite{5} analyze the response of the absorber only to one of the two
side components and do
not take into account their inevitable influence upon each other
(see, e.g., \cite{21}). As follows from the model of the CPO effect,
when the probe beam contains two components symmetric with
respect to the pump frequency, each of them contributes to the
other one and, in turn, experiences the influence of the latter to
its own amplitude and phase. Without considering this fact, the
suggested theoretical model is, in our opinion, inadequate. It is
also incorrect to apply the standard notion of group velocity to an
absorbing medium with a strong spectral dependence of
absorption (under these conditions, the spectrum and shape of
the light pulse should exhibit strong changes, and the notion of
group velocity needs to be redefined).

However, the main drawback of papers \cite{3,4,5,6} is that
the "dip" detected in the space of {\it difference} frequencies
(beat frequencies between "pump" and "probe") is
interpreted as a dip in the {\it optical} spectrum, which was not
really observed and, under conditions of the described
experiments, could not be observed. To detect the dip, spectral
width of the pump beam (as well as of the probe) should
be smaller than that of the dip. As applied to the case of
ruby crystal with a 37-Hz width of the "dip", the frequency of laser
light should be stable to within 14 (!) decimal digits (at least for
the times of the order of inverse width of the dip). In the
experiments \cite{3,4,5,6}, the widths of the laser sources evidently
exceeded the widths of the "burned" features by many orders of magnitude. Under
these conditions, position of the above "narrow dip", having no
preferences in a wide spectral range, becones physically
uncertain and loses any sense.

All the aforesaid is also true for other studies in which the
retarded dynamics of a saturable absorber
was attributed to the hole-burning effect under conditions of CPO
\cite{7,8}.

A different experimental situation was realized in \cite{22}, where as
the pump and the probe were used quasi-monochromatic beams of
independent laser sources, and the dip in the homogeneously
broadened spectrum of excitonic absorption of a semiconductor
structure with quantum wells was really detected in a classical
way as was implied in early theoretical papers on the CPO effect
(see, e.g., \cite{14}). Based on the results of the amplitude and phase
measurements, the authors concluded that the group velocity in
the system under study, due to a high steepness of the dispersion curve
in the region of the above dip, should be 9,600 m/s. This
conclusion is, in our opinion, not sufficiently well grounded. First of all, in
view of the already mentioned mutual influence of symmetric (with
respect to the pump frequency) spectral components of the light
pulse, the character of its interaction with the medium should
crucially depend on its phase structure. In particular, as has shown
our preliminary analysis, for a certain phase profile of the probe
pulse, it proves to be insensitive to the action of the pump
and incapable of detecting the dip. This fact indicates
again the nonlinear nature of the "hole" detected in the CPO effect and incorrectness of direct application of the Kramers-Kronig
relations to this spectral feature. And, lastly, as was
already mentioned, it is inconsistent to apply the notion of the
group velocity to the absorbing medium with a strong spectral
dependence of imaginary part of the refractive index.

\section{CONCLUSION}

The results of the above treatment allow us to conclude that the
experimental data of the papers \cite{3,4,5,6}, in which the "slow" or
"fast" light was allegedly observed under conditions of the
coherent population oscillations, can be easily interpreted in the
framework of the simplest model of saturable absorber. In our
opinion, the light pulse delay detected in these
experiments have nothing to do with the hole- (or anti-hole-) burning
in a homogeneously broadened absorption spectrum, as well as with
reduction of the group velocity of light in the
medium. These delays are
caused, in fact, by self-induced distortion of the pulse
propagating in the medium with intensity-dependent absorption. As for the
CPO-based hole-burning effect, it can be observed only under
conditions of quasi-monochromatic pump with its spectral width
smaller than the inverse population relaxation time. To observe
modification of the light group velocity in the medium under
condition of the coherent population oscillations, the probe
pulse spectrum should have approximately the same width. To
the present day, to the best of our knowledge, no experiment of this kind has been performed. One may assume
that, upon narrowing of the pump and probe pulse spectra, the
contribution to the pulse delay related to the refractive-index
dispersion feature
of the medium in the region of the pump frequency will become
noticeable on the background of the effects of retarded
absorption considered in this paper. However, taking into
account that this effect can be observed only in the
absorbing medium, and the time delay of the pulse may
constitute only a small fraction of its width \cite{9}, this "slow light"
will not differ from the "slow light" obtained using a saturable
absorber, and the prospects of its application for practical
purposes seem highly doubtful.

The authors are grateful to E.B.Aleksandrov for useful
discussions.

FIGURE CAPTIONS

Fig. 1. Simplified energy-level diagrams of Cr$^{3+}$ ion in the
ruby (a) and alexandrite (b) crystals. Under conditions of
sufficiently strong optical pumping in the $|a\rangle$ -- $|b\rangle$ channel, a
fast relaxation of electronic excitation from state $|b\rangle$ to metastable state
$|c\rangle$ redistributes populations of the $|a\rangle$ and $|c\rangle$ levels (within the
time intervals of the order of $\tau$) and changes the absorption
of the crystal at the wavelength of excitation.

Fig. 2. Time dependence of intensity of
the light beam transmitted through the usual (b) and inverse (c)
saturable absorber upon step-wise change of the incident light
intensity (a).

Fig. 3. Typical frequency dependences of the amplitude
$K(\omega)$, phase $\phi(\omega)$, and time delay $\Delta t(\omega)=\phi(\omega)/\omega$ of
intensity oscillations of the light beam transmitted through the
usual (a) and inverse (b) saturable absorber.

Fig. 4. Normalized light pulses at the entrance (solid lines) and
at the exit (dashed lines) of the usual (a) and inverse (b)
saturable absorber for several ratios of the pulse width $\delta$
and relaxation time $\tau$.  Calculations were performed for
the pulse propagating in the presence of a background
illumination (under conditions close to these of \cite{5}).

\begin{figure}
\epsfxsize=400pt
\epsffile{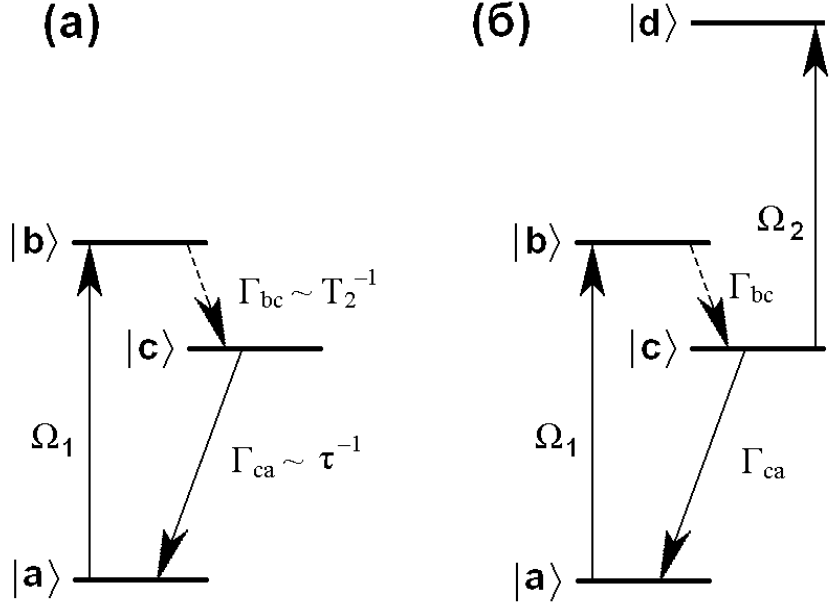}
\caption{Fig. 1. Simplified energy-level diagrams of Cr$^{3+}$ ion in the
ruby (a) and alexandrite (b) crystals. Under conditions of
sufficiently strong optical pumping in the $|a\rangle$ -- $|b\rangle$ channel, a
fast relaxation of electronic excitation from state $|b\rangle$ to metastable state
$|c\rangle$ redistributes populations of the $|a\rangle$ and $|c\rangle$ levels (within the
time intervals of the order of $\tau$) and changes the absorption
of the crystal at the wavelength of excitation.}
\end{figure}

\begin{figure}
\epsfxsize=400pt
\epsffile{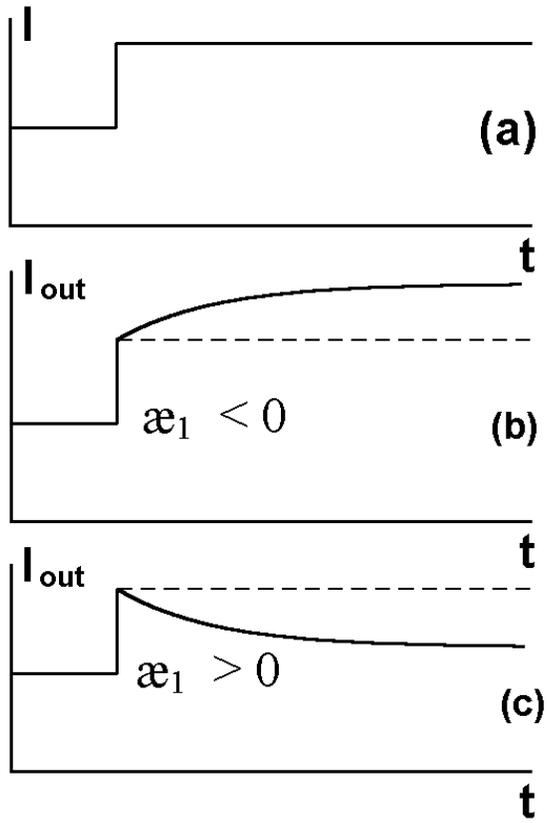}
\caption{Fig. 2. Time dependence of intensity of
the light beam transmitted through the usual (b) and inverse (c)
saturable absorber upon step-wise change of the incident light
intensity (a).}
\end{figure}

\begin{figure}
\epsfxsize=400pt
\epsffile{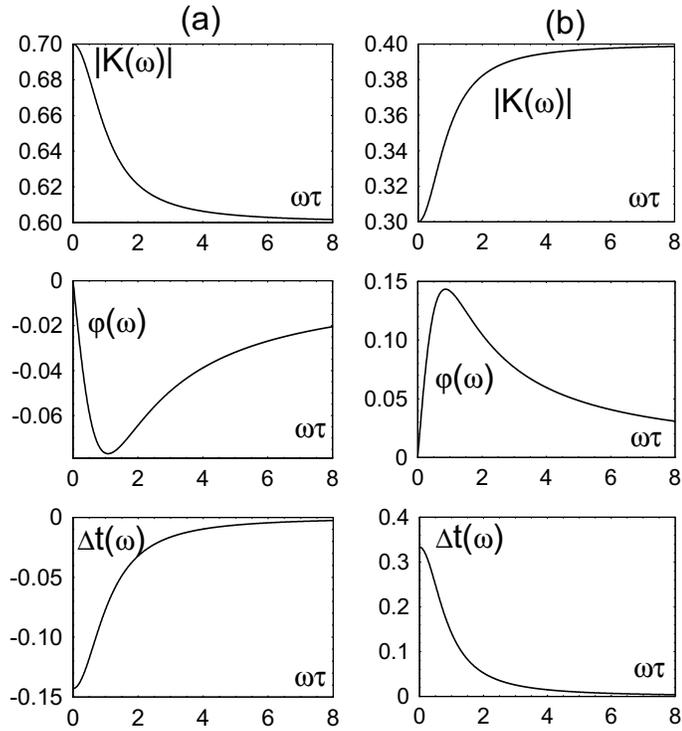}
\caption{Fig. 3. Typical frequency dependences of the amplitude
$K(\omega)$, phase $\phi(\omega)$, and time delay $\Delta t(\omega)=\phi(\omega)/\omega$ of
intensity oscillations of the light beam transmitted through the
usual (a) and inverse (b) saturable absorber.}
\end{figure}

\begin{figure}
\epsfxsize=400pt
\epsffile{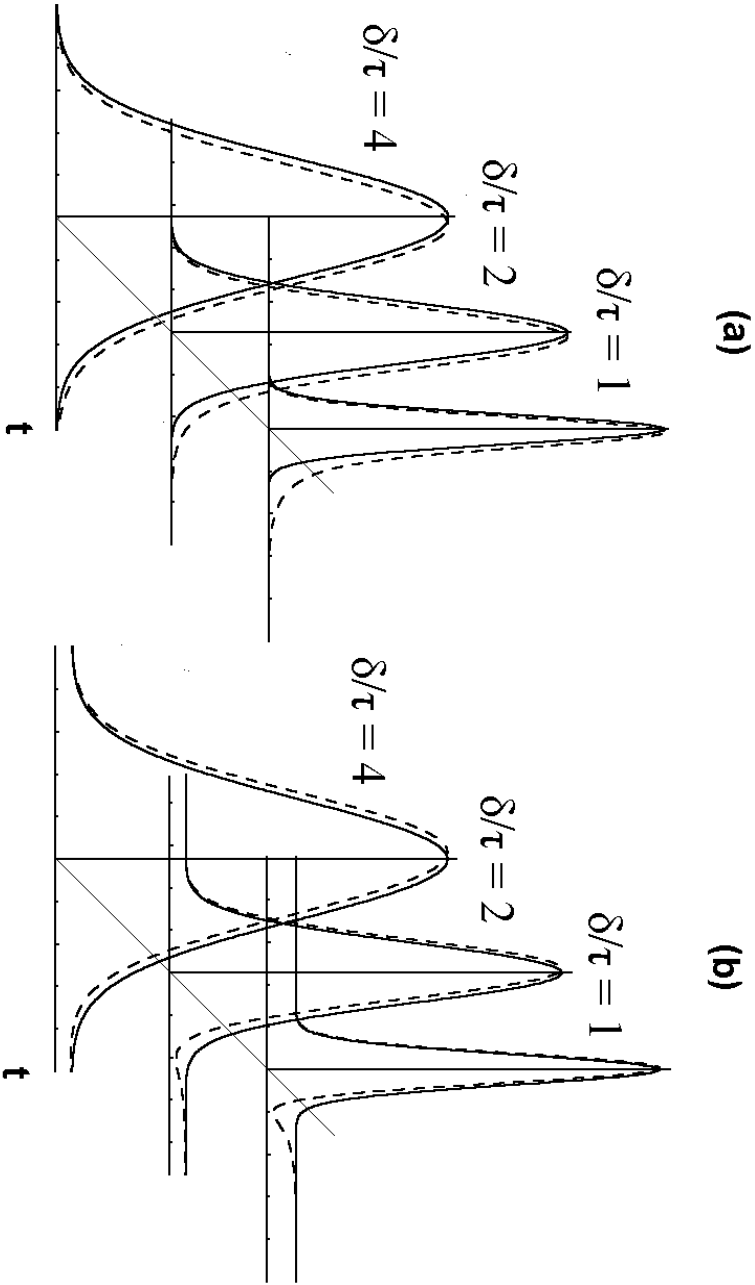}
\caption{Fig. 4. Normalized light pulses at the entrance (solid lines) and
at the exit (dashed lines) of the usual (a) and inverse (b)
saturable absorber for several ratios of the pulse width $\delta$
and relaxation time $\tau$.  Calculations were performed for
the pulse propagating in the presence of a background
illumination.}
\end{figure}

\end{document}